\documentclass[10pt,letterpaper]{article}
\usepackage{opex3}

\begin{document}

\title{Weak interference in the high-signal regime}

\author{Juan P. Torres$^{1,2,*}$, Graciana Puentes$^1$, Nathaniel Hermosa$^1$ and Luis Jose Salazar-Serrano$^1$}
\address{$^1$ICFO-Institut de Ciencies Fotoniques, UPC, Mediterranean
Technology Park, 08860 Castelldefels (Barcelona), Spain}

\address{$^2$Department of Signal Theory and Communications, Universitat
Polit{\`{e}}cnica de Catalunya, Castelldefels, 08860 Barcelona,
Spain}

\begin{abstract}
{\em Weak amplification} is a signal enhancement technique which
is used to measure tiny changes that otherwise cannot be
determined because of technical limitations. It is based on the
existence of a special type of interaction which couples a
property of a system, i.e., polarization or which-path
information, with a separate degree of freedom, i.e., transverse
position or frequency. Unfortunately, the weak amplification
process is generally accompanied by severe losses of the detected
signal, which limits the applicability of the {\em weak
amplification} concept. However, we will show here that since the
{\em weak measurement} concept is essentially an interference
phenomena, it should be possible to use the degree of interference
to get relevant information about the physical system under study
in a more general scenario, where the signal is not severely
depleted (high-signal regime). This can widen the range of systems
where the weak measurement concept can be applied. In this
scenario, which can be called generally {\em weak interference},
the idea of {\em weak value} does not convey any useful
information.
\end{abstract}


\email{juanp.torres@icfo.es}

\section{Introduction}
A weak measurement is a concept first introduced by Aharonov,
Albert, and Vaidman \cite{aharonov1988}, which describes a
situation where two subsystems interact, even though through a
weakly coupling process. Generally, the weak measurement idea is
presented in the general framework of Quantum Mechanics, where one
of the subsystems is assumed to be the {\em measuring device},
whose aim is to unveil the value of a property that characterizes
the other interacting subsystem ({\em the system}).

The fact that the coupling is weak can be seemingly
disadvantageous, since it is expected to produce an uncertainty in
the measurement larger than the values that should be
differentiated.  However, when appropriate initial and final
states of the system to be measured are selected, for instance by
choosing them to be nearly orthogonal, the mean value of the
reading of the measuring system can yield an unexpectedly large
value ({\em weak amplification}). Surprisingly, this value can lay
outside the range of small displacements of the measurement
pointer caused by each one of the possible states of the system.
Unfortunately, this is also accompanied by a severe depletion of
the intensity of the signal detected, due to the
quasi-orthogonality of the input and final states of the system.
This prevents the applicability of the weak measurement concept to
experiments which are already limited by a low signal-to-noise
ratio \cite{hosten2008}, since the intensity of the detected
signal is severely decreased.

For instance, when considering the refraction of a light beam in a
thin birefringent crystal, the two orthogonal linear polarizations
components of the optical beam are displaced a small distance
$\Delta$ compared with the beam waist \cite{ritchie1991}. For a
given initial $|\Psi_{in}\rangle$ and final $|\Psi_{out}\rangle$
states of the polarization of the system, so that $\langle
\Psi_{out}|\Psi_{in}\rangle \sim \epsilon$ ($\epsilon$ is small),
the mean value of the shift of the position of the light beam is
$\langle x \rangle/\Delta= \Re \left( \langle A
\rangle_{w}\right)$, where the weak value $\langle A \rangle_{w}$
is  defined as
\begin{equation}
\langle A \rangle _{w}=\frac{\langle \Psi_{out} |A
|\Psi_{in}\rangle}{\langle \Psi_{out} |\Psi_{in} \rangle} \sim
\frac{1}{\epsilon}.
\end{equation}
In this example, the power of the output signal ($P_{out}$) is
severely reduced, i.e., $P_{out}/P_{in} \sim \epsilon^2/2$. The
signal enhancement of the position of the beam due to weak
amplification can thus be observed only if the input signal
intensity ($P_{in}$) can be enhanced.

Most experimental realizations of weak measurements up to date
take place in this context. This is the case for experiments that
use the polarization of a light beam to reveal extremely small
spatial displacements \cite{hosten2008,ritchie1991}, and for
experiments that make use of the two counter-propagating paths in
a Sagnac interferometer to detect tiny beam deflections
\cite{ben_dixon2009}, or tiny frequency shifts
\cite{starling2010}. Interestingly, weak measurements can have a
wider range of applicability than originally conceived, appearing
naturally in the context of optical telecom networks
\cite{brunner2003}, and the description of the response function
of a system \cite{solli2004}.

Even though the idea of weak measurement was born in the context
of Quantum Mechanics, and is generally formulated in the language
of Quantum Mechanics, it can be understood as the consequence of
the constructive and destructive interference between the complex
amplitudes of different pointer states \cite{duck1989}. Each of
these correspond to the nearly equal readings of the measuring
device for each value of the state of the system. The concept
behind a weak measurement, also termed as {\em weak
amplification}, can thus be applied to any physical wave phenomena
and explained in classical terms as a wave interference process
\cite{howell2010}.

Here we will show that since a weak measurement is an interference
phenomenon, its effect should be noticeable even in the regime of
small losses, where the concept of {\em weak value} might not
convey any relevant information about the system or the measuring
device. This can open new applications based on weak interference,
especially for interactions where it is not possible to enhance
the signal-to-noise ratio of the measurement. For example, when
the intensity of the input signal cannot be increased.

\begin{figure}[t]
  \centering
  \includegraphics[scale=0.6]{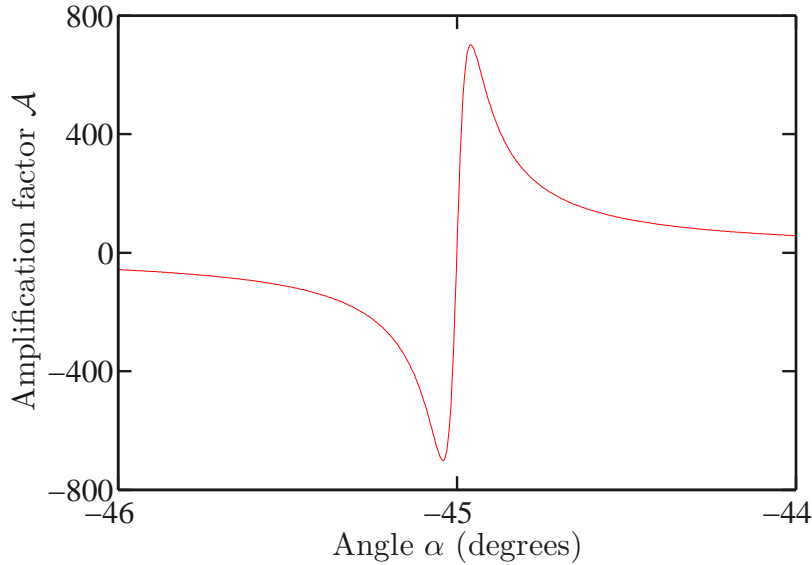}
  \caption{Weak amplification of the displacement $\Delta$ as a function of the
  output polarization of the system (angle $\alpha)$. The phase difference is
  $\theta=0.01^o$. We consider displacements $\Delta=\Delta_1=-\Delta_2=10$ nm, and the beam waist is $w_0=10 \mu$m.}
  \label{figure1}
\end{figure}

\section{Weak interference: general scenario}
For the sake of simplicity, let us consider a simple case which
shows the full potential of the scheme considered here. A linearly
polarized input light beam with polarization $|\Psi_{in}
\rangle=1/\sqrt{2}\, \left[ |H\rangle+|V\rangle\right]$ and
spatial shape $\Phi(x)$ propagates in a medium that couples the
polarization and spatial degrees of freedom, thereby generating a
polarization-dependent displacement of the photons, $\Delta_1$ and
$\Delta_2$.  The input state of the system (polarization) and the
measuring device (position of the beam) can be written, up to a
normalization constant, as
\begin{equation}
\label{input_state2} |in\rangle=|\Psi_{in} \rangle \otimes \int dx
\,\Phi(x) |x\rangle.
\end{equation}
The global state of the system and measuring device after the
interaction writes \cite{ritchie1991}
\begin{equation}
\label{output_state2} |out\rangle=\int dx \left\{
\Phi(x-\Delta_1)|H \rangle + \Phi(x-\Delta_2) \exp \left( i
\varphi\right) |V \rangle \right\} |x\rangle,
\end{equation}
where $\varphi$ stands for any polarization-dependent phase
difference that can occur during the interaction. Finally, before
being detected, the photons are projected into the polarization
state $|\Psi_{out}\rangle=\cos \alpha |H\rangle+\exp (i\xi)\sin
\alpha |V\rangle$. The intensity distribution of the output light
beam can now be written as
\begin{equation}
\label{spatial_distribution2} I(x)=\left| \cos \alpha
\,\Phi\left(x-\Delta_1\right)+\sin \alpha\,
\Phi\left(x-\Delta_2\right) \exp \left( i \theta \right)\right|^2,
\end{equation}
with $\theta=\varphi-\xi$.

Let us assume that the light beam is Gaussian with beam waist
$w_0$. Making use of the product theorem for Gaussian integrals
\cite{wolf1989}, the mean value of the position of the beam,
$\langle x \rangle=\int dx\,x\,I(x)/\int dx\, I(x)$ reads
\begin{equation}
\label{mean2} \langle x
\rangle=\frac{\Delta_+}{2}+\frac{\Delta_{-}}{2} \frac{\cos
2\alpha}{1+\gamma \sin 2\alpha \cos \theta},
\end{equation}
where
\begin{equation}
\label{gamma} \gamma=\exp \left[ -\frac{\left(
\Delta_1-\Delta_2\right)^2}{4w_0^2}\right],
\end{equation}
$\Delta_{+}=\Delta_1+\Delta_2$ and $\Delta_{-}=\Delta_1-\Delta_2$.
In all cases, $\Delta_1,\Delta_2 \ll w_0$, so $\gamma \sim 1$.

\begin{figure}[t]
  \centering
  \includegraphics[scale=0.6]{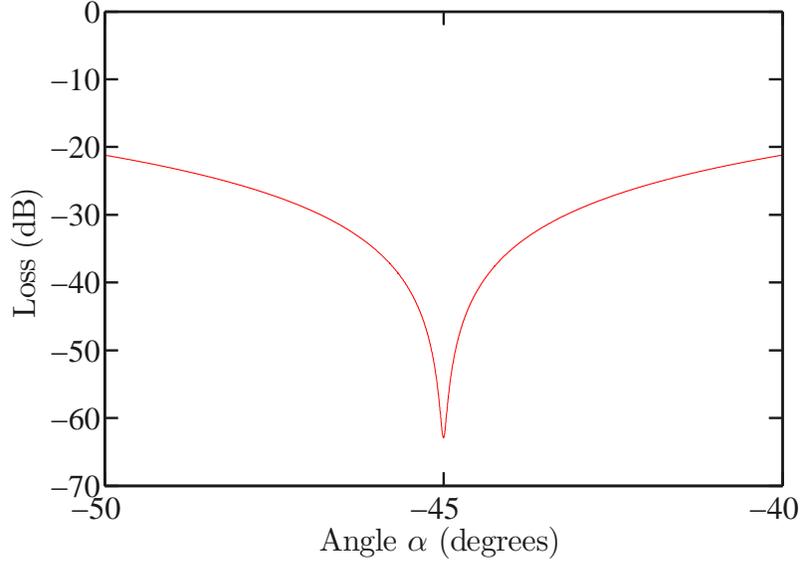}
  \caption{Signal loss as a function of the angle $\alpha$ of the
  polarization of the output state. The phase difference is $
\theta=0.01^o$, $\Delta_1=-\Delta_2=10$ nm and $w_0=10 \mu$m.}
  \label{figure2}
\end{figure}
Inspection of Eq. (\ref{mean2}) shows that the value of
$\Delta_{+}$ can not be amplified with the present scheme. In many
experiments \cite{hosten2008,ritchie1991}, $\Delta_1=-\Delta_2$,
so $\Delta_{+}=0$,  and as a consequence the weak measurement
amplifies all the relevant information. Therefore, the weak
amplification happens for $\Delta_{-}$, with an amplification
factor ${\cal A}$ that reads
\begin{equation}
\label{amplification} {\cal A}=\frac{\cos 2\alpha}{1+\gamma \sin
2\alpha \cos \theta}.
\end{equation}
Fig. \ref{figure1} shows the amplification factor for
$\theta=0.01^o$. The maximum amplification takes place for the
angle $\alpha_0=-1/2\,\sin^{-1} \left(\gamma \cos \theta \right)$,
where the factor reaches the value of ${\cal
A}_{max}=\left(1-\gamma^2 \cos^2 \theta \right)^{-1/2}$.

The loss of the system is given by
\begin{equation}
\label{loss} \frac{P_{out}}{P_{in}}=\frac{1}{2} \left[1+\gamma
\cos \theta \sin 2\alpha \right],
\end{equation}
where $P_{in,out}$ is the total input and output power of the
optical beam, when integrated over all space. Fig. \ref{figure2}
shows the loss in the measurement expressed in dB, as
$10\,\log(P_{out}/P_{in})$. Even though the enhancement of
$\Delta$ can be a large staggering value for angles close to
$-45^o$ (Fig. \ref{figure1} shows an enhancement close to $10^3$),
this is unfortunately accompanied by a severe loss penalty close
to $70$ dB (Fig. \ref{figure2}). For instance, if the goal of the
experiment is to attain a signal-to-noise ratio of 10 dB at the
measurement stage, the input signal has to be correspondingly
increased $70$ dB above this level.

\begin{figure}[t]
  \centering
  \includegraphics[scale=0.6]{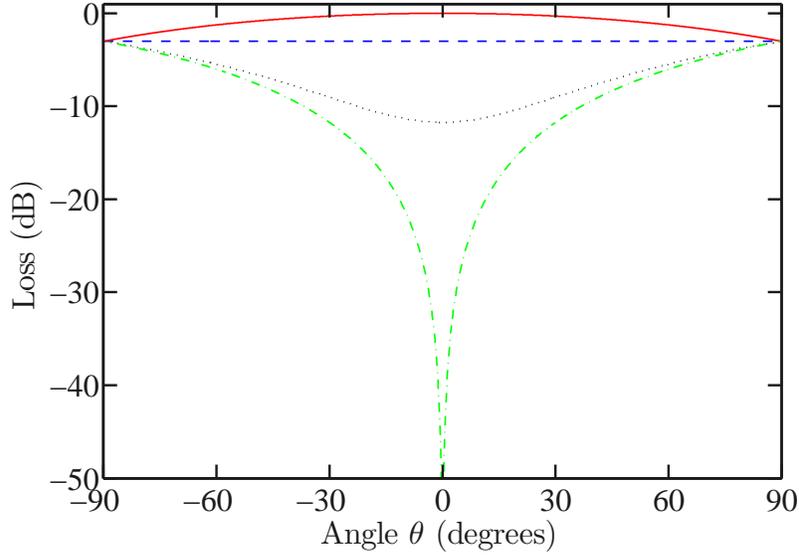}
  \caption{Signal loss as a function of the angle $\theta$.
  The beam waist is $w_0=10 \mu$m. Red solid line: $\alpha=45^o$; Blue dashed
line: $\alpha=0^o$; Black dotted line: $\alpha=-30^o$; Green
dotted dashed line: $\alpha=-45^o$.}
  \label{figure3}
\end{figure}

As it can be seen in Eq. (\ref{amplification}), the level of weak
amplification achievable depends on the value of $\theta$, which
should be chosen close to zero in order to achieve the maximum
amplification. This angle can be modified by choosing the
appropriate value $\xi$ of the polarization of the output state of
the system. The power of the output signal also depends on this
angle. Fig. (\ref{figure3}) shows the signal loss for a few
selected angles: $\alpha=-45^o$, $-30^o$, $0^o$ and $45^o$. In all
cases, the loss goes from $\log\left[\left(1+\sin 2\alpha
\right)/2\right]$ for $\theta=0$, to $-3$ dB for $\theta=\pm
\pi/2$, which corresponds to post-selecting polarizations $\cos
\alpha |H\rangle+\sin \alpha |V\rangle$ and $\cos \alpha
|H\rangle+i\sin \alpha |V\rangle$, respectively. Importantly, the
dependence of the losses on the polarization selected at the
output  for $\alpha \ne 0^o,90^o$ is a consequence of the
interference effect which is the essence of the weak measurement
concept \cite{duck1989}.

\section{Weak measurement in a high-signal regime}
\begin{figure}[t]
  \centering
  \includegraphics[scale=0.6]{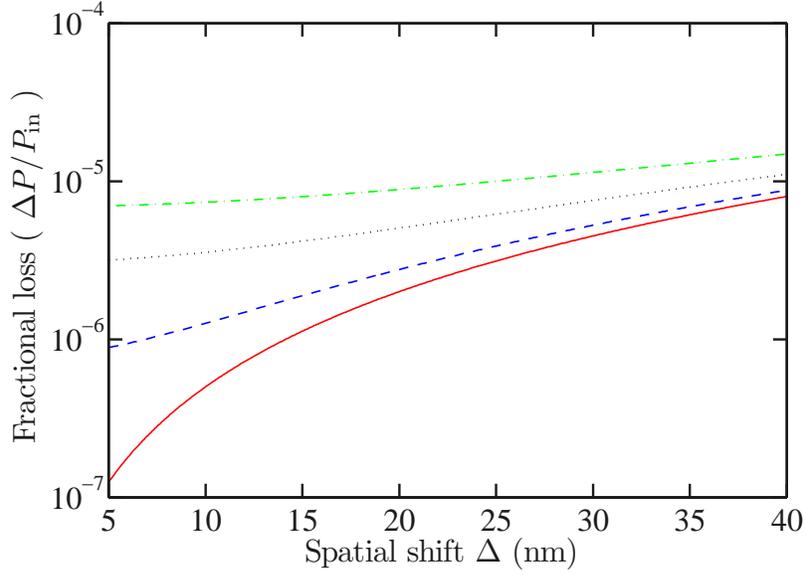}
  \caption{Fractional loss $\Delta P/P$ as a function of the spatial shift $\Delta$.
  $\alpha=45^o$ and $w_0=10 \mu$m. Red solid line: $\theta=0^o$; Blue dashed
line: $\theta=0.1^o$; Black dotted line: $\theta=0.2^o$; Green
dot-dashed line: $\theta=0.3^o$.}
  \label{figure4}
\end{figure}

In the low-signal regime, when the input and output polarization
states are quasi-orthogonal, the retrieval of information about
the value of $\Delta$ comes with a severe loss penalty. But since
the weak measurement is an interference phenomenon, we should be
able, in principle, to observe interference also in the
high-signal regime. In order to get information about the value of
$\Delta$, we can measure the fractional loss $\Delta P/P_{in}$,
where $\Delta P=P_{out}-P_{in}$. One obtains that
\begin{equation}
\label{ratio} \frac{\Delta P}{P_{in}}=\frac{1}{2} \left[\gamma
\cos \theta \sin 2\alpha -1\right].
\end{equation}
Figure \ref{figure4} shows the value of $\Delta P/P$ as a function
of the spatial shift $\Delta$ for $\alpha=45^o$ and $\theta=0$,
$0.1^o$, $0.2^o$ and $0.3^o$. The dependence of the fractional
loss on $\Delta$ comes from the relationship between $\gamma$ and
$\delta$, as given in Eq. (\ref{gamma}). Notice that in all cases,
the total losses of the system are below $3$ dB, which is
significantly below the loss found in the usual regime of weak
amplification, where losses can easily reach tens of dB for large
amplifications. Moreover, in all cases shown in Fig.
\ref{figure4}, the mean value of the beam position is $\langle x
\rangle=0$, so the weak value concept does not convey any relevant
information here.

The important point here is that by measuring the fractional loss,
one can determine the value of the displacement $\Delta$. The
maximum sensitivity of the scheme proposed is obtained for
$\alpha=45^o$ and $\theta=0^o$. By choosing other values for these
angles, one can decrease the fractional loss, making its detection
easier. However, this would decrease the sensitivity, making the
distinction between different spatial shifts $\Delta$ more
difficult. Note that here we are assuming that there are not other
sources of polarization-dependent losses.

What is, in general, the minimum fractional loss measurable? In
\cite{freudiger2008,saar2010}, a high-frequency detection scheme
for the detection of Raman gain in Stimulated Raman Scattering
process was able to detect a fractional loss of the order of
$\Delta P/P < 10^{-7}$. The key point is to modulate the input
signal at MHz rates to remove the low-frequency laser noise,
implementing a detection scheme that is effectively shot-noise
limited.

\section{Conclusions}
In conclusion, we have shown that since the idea of weak
measurements is indeed an interference phenomenon, it is possible
to obtain relevant information about the weak interaction of the
physical system under study even in a regime where the signal is
not depleted, and the concept of {\em weak value} does not convey
any useful information. We demonstrate that we can measure tiny
quantities generated during a weak interaction by detecting a
measurable interference effect in the high-signal regime, as
opposed to the usual case of weak amplification, which suffers
from severe losses. Therefore, this widens the applicability of
the weak measurement concept, allowing its use in a broader range
of systems.

Finally, we should mention that even though we have discussed our
ideas in the context of the specific case of
polarization-dependent spatial shifts of optical beams, the main
conclusions applies also to other systems, as well as to other
degrees of freedom, i.e., frequency. Our scheme would also apply
to systems in a higher-dimensional space, so that the total input
state writes
\begin{equation}
\label{input_state3} |in\rangle=\frac{1}{\sqrt{n}} \sum_{i=1}^{n}
|u_i \rangle \otimes \int dx\, \Phi(x) |x\rangle,
\end{equation}
where $|u_i \rangle$ is a base of the N-dimensional space.

An example of this kind of interaction, where our system could be
implemented, is the rotational Doppler frequency shift imparted to
a beam with orbital angular momentum (OAM) when it traverses a
rotating optical device which introduces a time-varying
OAM-dependent phase shift, such as a rotating Dove prism. In this
case, the spatial and frequency degrees of freedom are coupled
\cite{courtial1998,vasnetsov2003}, so that the state of the light
beam after traversing the rotating Dove prism is
\begin{equation} \label{output_state4}
|out\rangle=\sum_m \int d\omega  \Phi(\omega+2m\Omega)|m\rangle
|\omega\rangle
\end{equation}
Now, the interaction introduces frequency shifts
$\Delta_m=2m\Omega$, where $m$ is the OAM mode index.

\section*{Acknowledgments}
This work was supported by the Government of Spain
(FIS2010-14831), by the European project PHORBITECH (FET-Open
grant number: 255914), and by Fundacio Privada Cellex Barcelona.
GP acknowledges financial support from Marie Curie International
Incoming Fellowship COFUND.

\end{document}